\documentclass[preprint,journal]{vgtc}       

\ifpdf
  \pdfoutput=1\relax                   
  \usepackage{graphicx}                
  \DeclareGraphicsExtensions{.pdf,.png,.jpg,.jpeg} 
\else
  \usepackage{graphicx}                
  \DeclareGraphicsExtensions{.eps}     
\fi%

\graphicspath{{figures/}{pictures/}{images/}{./}} 

\usepackage{microtype}                 
\PassOptionsToPackage{warn}{textcomp}  
\usepackage{textcomp}                  
\usepackage{mathptmx}                  
\usepackage{times}                     
\usepackage{cite}                      
\usepackage{tabu}                      
\usepackage{booktabs}                  
\usepackage[pdflang={en-US},pdftex]{hyperref}
\usepackage{color}
\usepackage{tabularx}
\usepackage[table]{xcolor}
\usepackage{ragged2e}
\usepackage{dblfloatfix}    
\usepackage{wasysym}
\usepackage{lettrine}
\usepackage{enumitem}
\usepackage{microtype}
\usepackage[export]{adjustbox}
\usepackage{graphicx} 
\usepackage{animate}

\ieeedoi{10.1109/TVCG.2019.2934397}

\onlineid{0}

\vgtccategory{Research}
\vgtcpapertype{please specify}

\title{A Comparative Evaluation of Animation and Small Multiples\\~for Trend Visualization on Mobile Phones}

\author{Matthew Brehmer, Bongshin Lee, Petra Isenberg, and Eun Kyoung Choe}
\authorfooter{
\item
 \revision{Matthew Brehmer is an independent researcher; he conducted this work while with Microsoft Research. E-mail: mb@mattbrehmer.ca.}
\item
 Bongshin Lee is with Microsoft Research. E-mail: bongshin@microsoft.com.
\item
 Petra Isenberg is with Inria. E-mail: petra.isenberg@inria.fr.
\item
 Eun Kyoung Choe is with the University of Maryland, College Park. E-mail: choe@umd.edu.
}

\shortauthortitle{Brehmer \MakeLowercase{\textit{et al.}}: Too Small for Multiples?}

\abstract{We compare the efficacy of animated and small multiples variants of scatterplots on mobile phones for comparing trends in multivariate datasets. Visualization is increasingly prevalent in mobile applications and mobile-first websites, yet there is little prior visualization research dedicated to small displays. In this paper, we build upon previous experimental research carried out on larger displays that assessed animated and non-animated variants of scatterplots. Incorporating similar experimental stimuli and tasks, we conducted an experiment where 96 crowdworker participants performed nine trend comparison tasks using their mobile phones. We found that those using a small multiples design consistently completed tasks in less time, albeit with slightly less confidence than those using an animated design. The accuracy results were more task-dependent, and we further interpret our results according to the characteristics of the individual tasks, with a specific focus on the trajectories of target and distractor data items in each task. We identify cases that appear to favor either animation or small multiples, providing new questions for further experimental research and implications for visualization design on mobile devices. Lastly, we provide a reflection on our evaluation methodology. 
}


\teaser{
  \centering
  \animategraphics[height=3in,loop,autoplay]{10}{figures/task4/task4_}{008}{134}
  \includegraphics[height=3in]{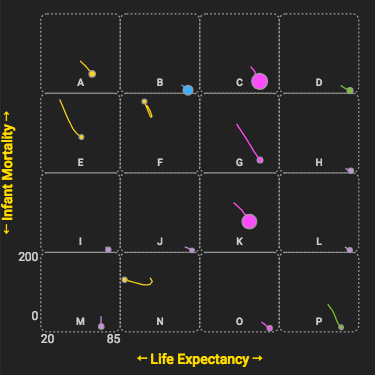}
  \caption{Cropped screen captures from our experimental mobile application. Left: an instance of the animation condition (\textbf{open this PDF in Acrobat Reader to view the animation}). Right: an instance of the small multiples condition, \revision{in which the position of each item reflects its values in the year 2000; a ``trail'' encodes the item's previous positions from 1975 to 1999}.} 
  \label{fig:teaser}
}

\vgtcinsertpkg

\newcommand{\bstart}[1]{\vspace{1mm} \noindent{\textbf{#1:}}}

\newcommand{\bpstart}[1]{\vspace{1mm} \noindent{\textbf{#1.}}}

\definecolor{highlight}{RGB}{139,0,0}
\definecolor{revision}{RGB}{217,95,2}
\definecolor{bongshin}{RGB}{102,166,30}
\definecolor{petra}{RGB}{217,95,2}
\definecolor{eunkyoung}{RGB}{117,112,179}

\definecolor{animation}{RGB}{31,119,180}
\definecolor{multiples}{RGB}{255,127,14}
\definecolor{diff}{RGB}{76,146,76}

\newcommand{\animation}[1]{\textcolor{animation}{\textbf{#1}}}
\newcommand{\multiples}[1]{\textcolor{multiples}{\textbf{#1}}}
\newcommand{\diff}[1]{\textcolor{diff}{\textbf{#1}}}

\newcommand{\highlight}[1]{\textcolor{highlight}{#1}}

\newcommand{\revision}[1]{{#1}}

\newcommand{\etal}{et al.}
\newcommand{\eg}{e.g.,~}
\newcommand{\ie}{i.e.,~}

\newcommand{\inlinevisImproved}[3]{\raisebox{#1}[0pt][0pt]{\includegraphics[height=#2]{#3}}}

\newcommand*{\indicator}{\inlinevisImproved{-2pt}{1.1em}{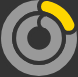}} 

\newcommand*{\indicatorb}{\inlinevisImproved{-2pt}{1.1em}{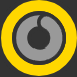}}

\newcommand*{\indicatorc}{\inlinevisImproved{-2pt}{1.1em}{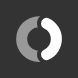}}

\begin{document}

\firstsection{Introduction}
\label{sec:introduction}

\maketitle

Visualizing data on mobile phones is increasingly prevalent in practice, yet it has attracted little attention from the visualization research community~\cite{Lee2018b}.
Many mobile applications now incorporate visualizations~\cite{Ros2014,Sadowski2015}, and there is a growing demand for responsive and mobile-first visualization design~\cite{Hinderman2016}, particularly in the context of news graphics~\cite{Aisch2016}.
Phones now have powerful processors and high-resolution displays that can accommodate a large number of graphical elements with a fine level of detail.
However, we lack guidance on how to design visualizations for mobile phones. 
Spanning the contexts of news consumption and informal data analysis, our work specifically concerns scenarios in which people encounter visual representations of quantitative multivariate data on mobile phones, whereupon they may identify and compare trends in the data. 
Animation and small multiples are two common design choices for facilitating these tasks. 


\begin{figure*}[b!]
    \vspace{-0.3cm}
    \includegraphics[width=\linewidth]{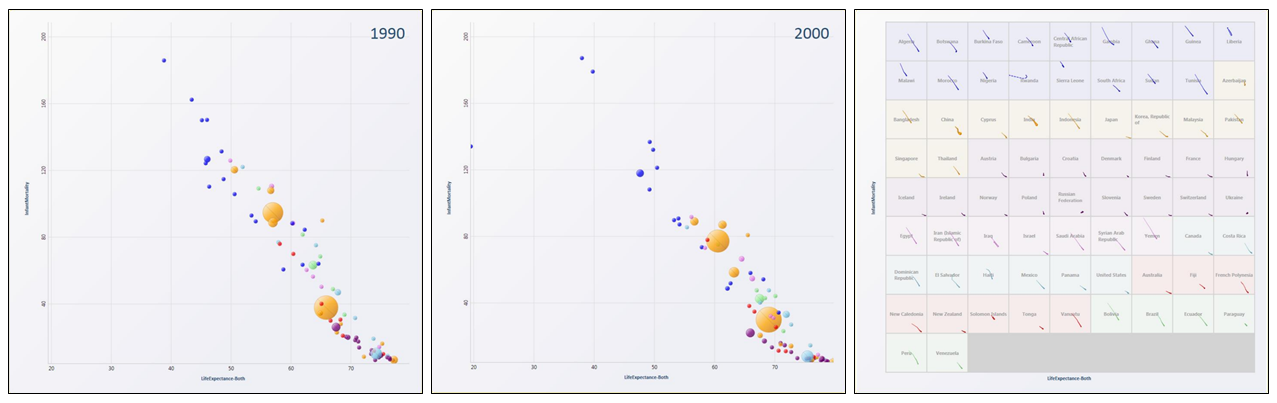}
    \vspace{-0.5cm}
    \caption{Trend visualization designs considered in Robertson~\etal's 2008 study~\cite{Robertson2008}. The left and middle frames are examples from their animation condition, while the right frame is an example from their small multiples condition. We incorporate a subset of their data and tasks in our experiment, albeit with a smaller dataset of 16 nations (cf. \autoref{fig:teaser}); \revision{we revisit the question of visualizing larger datasets on mobile displays in \autoref{sec:discussion:extensions}.}}
    \label{fig:robertson}
\end{figure*}

Hans Rosling famously used animated scatterplots to illustrate trends in his influential TED conference presentations~\cite{Rosling2006,Rosling2007}.
Shortly thereafter, Robertson~\etal~\cite{Robertson2008} conducted an experiment devoted to trend comparison tasks. 
They found that animated scatterplots were associated with lower task performance relative to a static alternative, a small multiples grid where the trajectory of each item was plotted as a `trace' within each grid cell.
Despite their findings, we continue to see animated charts in practice.
Furthermore, Robertson~\etal's study was conducted using large displays, and we questioned whether small multiples would remain to be a viable design choice for small displays. 
We therefore conducted a study to assess the efficacy of animation and small multiples in a mobile-only experiment with 96 crowdworkers.

We found that 
participants using the small multiples variant completed seven of the nine tasks in less time than those using animation; they did so without sacrificing accuracy, except in two tasks. However, those using small multiples reported feeling slightly less confident in their responses than those using animation. 
To further interpret our results, we considered how the characteristics of each trend comparison task affect performance, such as the number of correct responses and the trajectories of target items.
Our interpretations lead to new hypotheses about the efficacy of animation and small multiples for trend comparison tasks on mobile phones: 
trajectory reversals and direction-of-motion outliers may be more apparent with animation, while prominent trajectory lengths and angles may be more apparent with small multiples.
Finally, we offer a reflection on our evaluation methodology, as we hope to motivate future research addressing the challenges of mobile-specific visualization.  


\begin{figure*}[b!]
    \vspace{-0.2cm}
    \includegraphics[width=\linewidth]{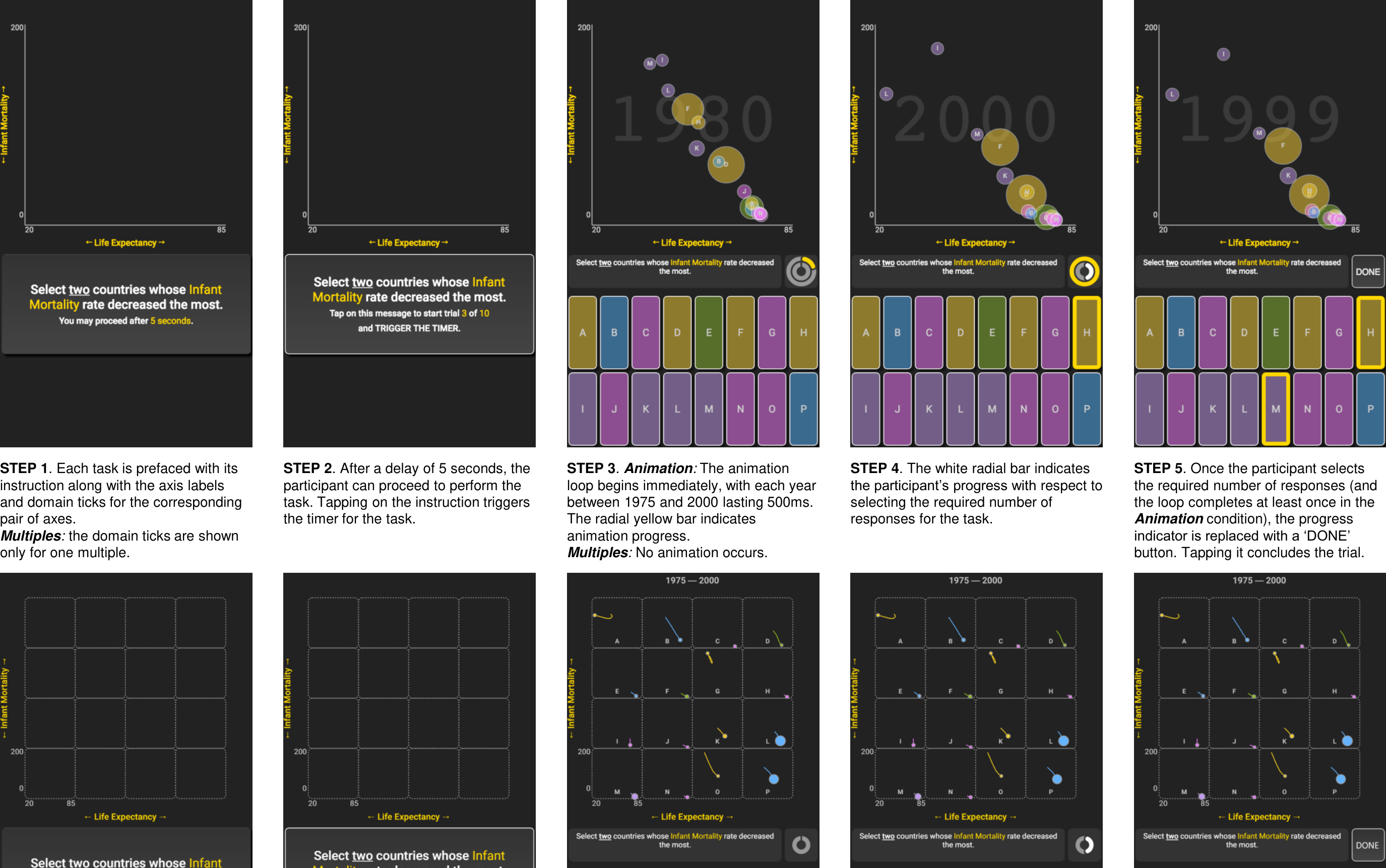}
    \vspace{-0.25cm}
    \caption{The five steps of a task (Task 5) in the Animation condition (top) and in the Multiples condition (bottom), comprised of a instruction reading phase (Steps 1--2) and a task completion phase (Steps 3--5). The alphabetical response interface was identical across the two conditions.}
    \label{fig:task_example}
\end{figure*}

\section{Related Work}
\label{sec:background}


We situate our experiment in relation to several areas of visualization research and practice, touching upon the topics of mobile data visualization, trend visualization, and evaluation. 

\subsection{Visualization on Mobile Devices}
\label{sec:background:mobile}

Over the past decade, we have seen a growing interest in visualization for mobile devices~\cite{Lee2018b,Watson2015}. 
Responsive~\cite{Andrews2018,Hinderman2016} and {\it mobile-first} visualization designs have become integral to the production of news graphics~\cite{Aisch2016,Hurt2015,Sam2018}, which are now predominantly consumed from a mobile device, including from dedicated news reader applications, mobile browsers, and social media applications.
Visualization is also now prevalent in mobile applications~\cite{Sadowski2015} \revision{and particularly those relating to health, activity tracking, and finance}.

While practitioners have catalogued examples and design patterns for visualization on mobile devices~\cite{Ros2014,Sadowski2015}, a need remains for best practices grounded in empirical evidence with regards to mobile-specific visualization design~\cite{Blumenstein2016}. 
Existing research has thus far considered various datatypes, such as continuous univariate time-series data~\cite{Chen2017}, hierarchical data \cite{AlTarawneh2015}, and range data~\cite{Brehmer2019}, as well as different application contexts, such as healthcare~\cite{Chittaro2006} and public transit navigation~\cite{Kay2016}.
We add to the existing body of work considering multivariate time-series, which is common in economics, public health, and business intelligence. 
Our data is from a public repository of international economic and public health indicators nations~\cite{UNStats}.

Other recent research has considered novel interaction design choices for visualization on tablet and mobile devices (\eg~\cite{Baur2012,Drucker2013,Jo2015,Jo2017,Schwab2019}).
While we recognize the importance of identifying suitable mobile interaction design choices, we focus here on visual perception, comparing a static small multiples design with a non-interactive animated design.

\subsection{Visualizing Trends with Animation and Small Multiples}
\label{sec:background:trends}

Our work focuses on trend visualization with multivariate datasets, wherein each item in a dataset has multiple attributes that change over time. Animated charts depicting trends have become increasingly common in recent years, particularly in a news media context and encompassing a variety of datatypes and visual encoding design choices~\cite{Groeger2017,SingerVine2017}. They are designed to be consumed casually without any interaction~\cite{Groeger2015}, as many will be consuming this content from a mobile device.
The perceived efficacy of animation has led to an ``animation on the phone, small multiples on the desktop'' design pattern for news graphics~\cite{Boyer2015} (\eg~\cite{NPRWalmart}), in which two versions of a graphic are produced, one for desktop and another for mobile.  Meanwhile, we have also observed the prevalence of small multiples designs in mobile applications and activity-tracking applications in particular (\eg~the iOS Activity app~\cite{Activity2018}).
Ultimately, seeing the use of both small multiples and animation on mobile phones in practice prompted us to the perform the study described below. 

Perhaps the most iconic examples of trend visualization are those featured in Hans Rosling's TED conference presentations~\cite{Rosling2006,Rosling2007}. 
Inspired by highly enthusiastic responses from the viewers of these presentations,
Robertson~\etal~\cite{Robertson2008} investigated the effectiveness of animation in trend visualization.
They compared Rosling's animation approach to trend visualization against a static small multiples variant and a static superimposed ``trails'' variant with small and large datasets (18 and 80 data points, respectively), both in simulated presentation and analysis settings.
In the simulated presentation setting, which tried to mimic the experience of Rosling's TED talks, participants listened to a scripted description of trends in the data. 
\autoref{fig:robertson} shows examples of their animated and small multiples designs. 
The study results showed that people using a small multiples design completed trend comparison tasks in an analysis context more quickly and more accurately than those using an animated design. 
However, people found animation to be more fun and exciting, and animation was associated with faster task completion times in the presentation context.

Robertson~\etal's study was considerably impactful, receiving a 10-year ``Test of Time'' award at IEEE InfoVis 2018. 
The study inspired researchers to make similar comparisons between animation and small multiples in the context of other datatypes, such as dynamic networks~\cite{Archambault2011} and flow maps~\cite{Boyandin2012}.
We also draw inspiration from Robertson~\etal's study and incorporate a subset of their experimental stimuli and tasks.
However, our study should be seen as a sequel and not a reproduction.
We compare the relative merits of animation and small multiples \revision{for a small dataset of 16 items} in a mobile device context.
We also excluded the presentation setting featuring a scripted aural description of trends.
In addition, we focus on individual task characteristics in our analysis because previous research suggests that the advantages of animation and small multiples in terms of speed, accuracy, and subjective preference may be task-dependent.

\subsection{\revision{(Mobile)} Visualization Evaluation Studies}
\label{sec:background:perception}

In response to a call for more empirical evidence to guide the design of visualization for mobile devices~\cite{Blumenstein2016}, researchers have recently started to take experimental approaches for mobile visualization~\cite{Blascheck2019,Brehmer2019,Schwab2019}.
Our experimental comparison of visualization design alternatives continues a tradition established by Cleveland and McGill~\cite{Cleveland1984} and continued by the crowdsourced graphical perception work of Heer and Bostock~\cite{Heer2010}. 
Our experiment also makes use of a crowdsourcing platform, which serves to circumvent some of the limitations of directly observed lab studies while providing a diverse and large participant pool~\cite{Borgo2017,Borgo2018}. 
\revision{Our work follows two recent crowdsourced experiments relating to visualization on mobile phones: our previous study comparing alternative ways to visualize ranges on mobile phones~\cite{Brehmer2019}, and Schwab~\etal's study of panning and zooming techniques with temporal data~\cite{Schwab2019}.
Our experiment adopts a similar protocol to the former, though in addition to different stimuli and tasks, we employed new techniques to improve the quality of responses.}
We will revisit the prospects for a crowd-based experimentation methodology for assessing visualization design choices on mobile phones later in \autoref{sec:discussion:method}. 
\section{Experiment}
\label{sec:experiment}

We designed and conducted a crowdsourced experiment on mobile phones to compare the efficacy of animated and small multiples variants of scatterplots. 
Participants performed trend comparison tasks where the values of data items varied across multiple dimensions over time.


Robertson~\etal~\cite{Robertson2008} indicated that people are likely to complete trend comparison tasks in less time and with a higher degree of accuracy when using small multiples scatterplots relative to when using animated scatterplots. 
As their experiment was performed using a 21-inch monitor, we were inspired to study how people perform trend comparison tasks on mobile phones: 
we wanted to see if small multiples remain to be a viable design choice in this context, and to quantify differences in performance between animation and small multiples. 

We also planned to examine performance across tasks having varying characteristics, as this would allow us to form new expectations with regards to how these characteristics contribute to performance differences. 
These characteristics include the initial and final distributions of items and the trajectories of target and distractor items over time, as indicated in \autoref{fig:tasks}.

\subsection{Multivariate Trend Data}
\label{sec:experiment:data}

To maximize external validity, we opted to use real data as opposed to synthetic data in our experiment.
Similar to Hans Rosling's TED conference presentations~\cite{Rosling2006,Rosling2007} and by Robertson~\etal's study~\cite{Robertson2008}, our stimuli data included economic and public health indicators for 16 nations over time, which was retrieved from the United Nations Common Database~\cite{UNStats}. 

The data in our experimental stimuli included one categorical dimension corresponding to the nation's region \revision{(\ie continent)} and nine quantitative dimensions corresponding to various indicators.
Each nation has a set of indicator values for each year between 1975 and 2000.
The indicators were population, GDP per capita, energy consumption, life expectancy (men), life expectancy (women), life expectancy (both sexes), infant mortality, the number of personal computers, and arable area.
We anonymized the nations' names, randomly assigning letters A--P to each nation for each participant. 
\revision{We did this to ensure that any preconceived beliefs regarding the economic or public health conditions of various nations did not factor into participants' responses.}




\subsection{Two Conditions: Animation vs. Small Multiples}
\label{sec:experiment:layout}

Our experimental conditions involve two variants of scatterplots for displaying multidimensional data over time: an \textbf{Animation} condition and a Small \textbf{Multiples} condition (Figures~\ref{fig:teaser} and~\ref{fig:task_example}). 
Both variants featured a mapping of two quantitative dimensions to the {\tt x} and {\tt y} positions of the points as well as a mapping of a third quantitative dimension to the size of the points. 
We varied the {\tt x} and {\tt y} mappings throughout our experiment, while the size of the marks was consistently mapped to the corresponding nations' populations. 
We mapped the color of each point to the nation's region. We selected and evaluated our color palette using Meeks and Lu's Viz Palette tool~\cite{Meeks2018}.
The size~\textrightarrow~population and color~\textrightarrow~region mappings are conventional  
and the default settings of Gapminder's ``Bubble Chart'' tool~\cite{Gapminder}. 


\bstart{Animation}
In this condition, each point's {\tt x}, {\tt y}, and size properties interpolate from one year to the next, with 500ms elapsing between years, resulting in a 12.5 second-long animation. \revision{We include a 10 FPS example of the animation condition in \autoref{fig:teaser}} (this figure animates when this document is viewed in Acrobat Reader). 
The supplemental video also presents several instances of the animation condition.
Finally, we annotated each item with the corresponding nation's letter (A--P).

We did not provide any interactive control over the animation; participants could not scrub along a timeline or jump to a later or earlier frame in the animation.
We consider the presence and absence of interactivity to be a separate experimental factor, and we expect that participants would interact with animation controls to varying extents, thereby confounding our analysis. 
Instead, we draw inspiration from non-interactive animated news graphics~\cite{Groeger2017} in that our animation plays in an automated continuous loop between 1975 and 2000, with a yellow radial bar to the bottom right of the chart to indicate the progress of the animation:  \indicator~\textrightarrow~\indicatorb\hspace{1mm}(see \autoref{fig:task_example}-top, Steps 3 and 4).


\bstart{Multiples}
In this condition, we divide the display into a 4x4 grid, with each grid cell allocated to one of the 16 nations in the dataset, ordered alphabetically from nation A to nation P and annotated accordingly. 
All of the grid cells share the same quantitative domains for the points' {\tt x} and {\tt y} positions. 
\revision{The position of each item reflects its values in the year 2000}; emanating from this point is a ``trail,'' a path element traversing the {\tt x-y} coordinate space, encoding the point's previous positions from 1975 to 1999 (\autoref{fig:task_example}-bottom), forming a connected scatterplot~\cite{Haroz2016}. 



\subsection{Tasks}
\label{sec:experiment:task}

Each of our tasks posed a multiple choice question about changes in economic and public health indicator values over 25 years.
These tasks were adapted from those used in Robertson~\etal's study.
We asked participants to select one, two, or three nations that exhibit a described behavior from an alphabetical list of nations shown below the chart (\autoref{fig:task_example}, Steps 3--5). 
Given this format, it was possible for participants to have partially correct responses to the tasks. 
For tasks with more than one required response, a white radial bar to the bottom right of the chart indicated one's progress with respect to completing the task: \indicatorc\hspace{1mm}(see \autoref{fig:task_example}, Step 4).
Once a participant selected the required number of responses, the progress indicator was replaced with a `Done' button (\autoref{fig:task_example}, Step 5) and they could not make additional selections. They could, however, change existing selections by de-selecting one item and then selecting another.

After piloting to get a sense of the relative difficulty of the tasks, we set aside three of the easiest ones as training tasks. 
For each training task, we showed participants feedback indicating the correctness of their response; if incorrect, they could try again until correct, and we highlighted the correct response after two failed attempts.

\begin{figure}[tb]
    \includegraphics[width=\linewidth]{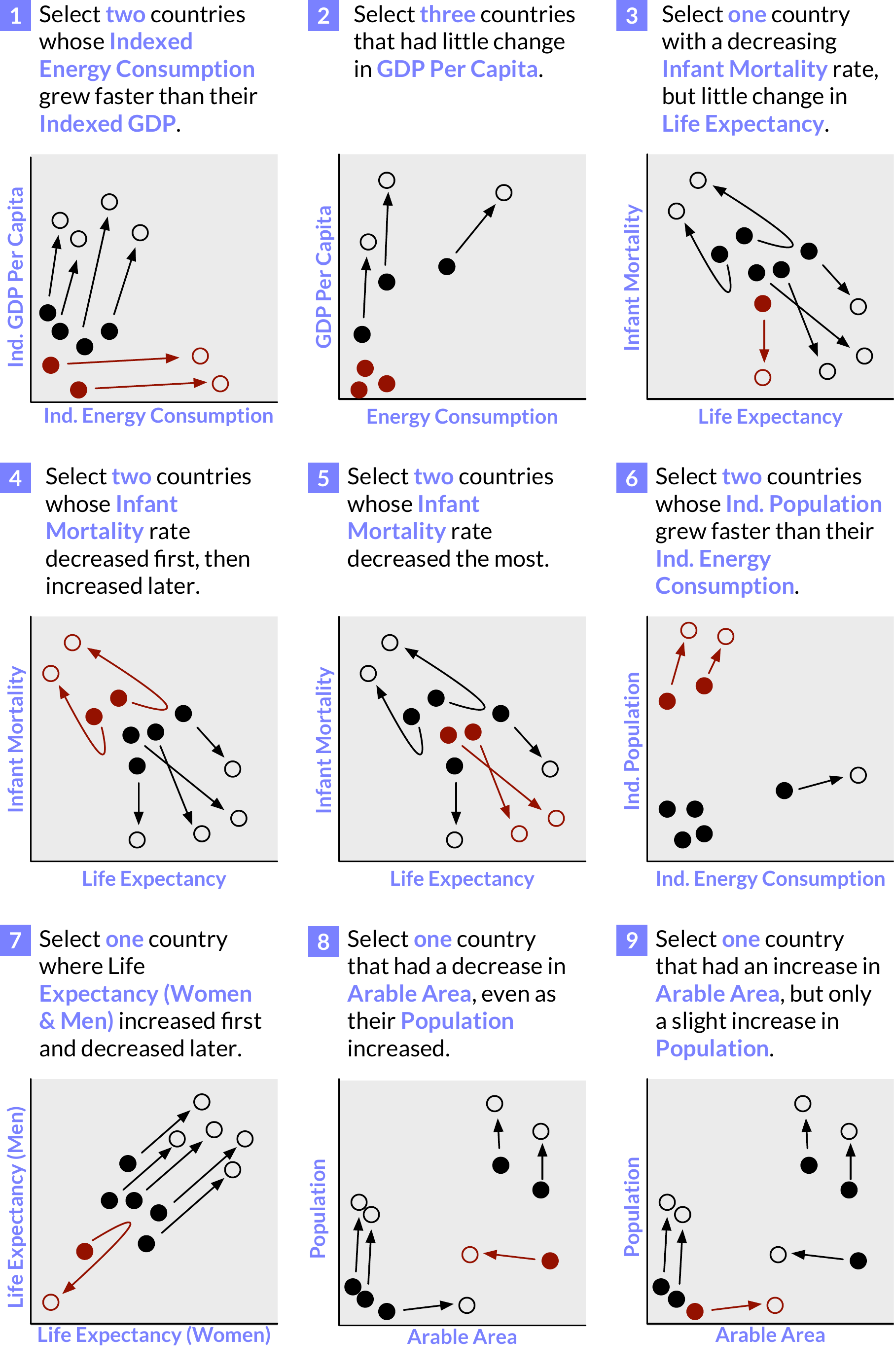}
    \vspace{-0.5cm}
    \caption{Diagrams of the nine tasks featured in our experiment. The starting and ending positions of \textbf{\highlight{target}} and \textbf{distractor} items are indicated by filled and unfilled circles, respectively. Note that the number of distractor items is reduced for clarity of illustration, as the total number of (target + distractor) items was 16 in every task. Similarly, the items are shown with uniform size for clarity of illustration, whereas the size of items was mapped to the corresponding nation's population in each task. These diagrams are preceded by the instruction shown to participants.}
    \label{fig:tasks}
    \vspace{-0.5cm}
\end{figure}

The nine non-training tasks are illustrated in \autoref{fig:tasks}, and these diagrams are preceded by the corresponding task instruction shown to participants. 
The task diagrams indicate the starting and ending positions of \textbf{\highlight{target}} and representative \textbf{distractor} items, as well as how targets and distractors exhibit different behavior across the nine tasks.

\subsection{Experiment Design}
\label{sec:experiment:design}

We conducted a between-subjects online experiment, where each participant experienced either the Animation or the Multiples condition.
Altogether, the experiment required approximately 10 minutes to complete, which included an introductory tutorial, three training tasks, nine timed tasks, one quality control task, and a concluding survey featuring subjective response questions.
For each experimental session, we randomized the assignment of nations to letters A through P as well as the assignment of colors to regions.
We also shuffled the order of task presentation such that no two consecutive tasks shared the same dimensional mapping to either the {\tt x} or {\tt y} axis.
We presented each participant with one of five shuffled task orderings, as this allowed us to identify potential effects of task ordering on performance.

The tutorial introduced participants to scatterplots and to either the Animation or the Multiples variant. 
It also introduced participants to the steps of a task (\autoref{fig:task_example}) and the required response interactions.  

Between two of the timed tasks in the shuffled task ordering, we inserted a quality control task to test the participants' attention and essentially their ability to interpret a scatterplot.
Specifically, this quality control task asked participants to ``Select the two countries having the largest Population in the year 2000.''
For this task, we redundantly encoded countries' populations to both the {\tt x}-position and the size of their corresponding points; furthermore, this task did not require any judgment of change over time, as the two countries corresponding to the correct responses had the largest population irrespective of the year.   
When a participant responded incorrectly to this quality control task, we used this as an indicator of inattention or as an inability to correctly interpret a scatterplot, and we excluded the participant's data from any subsequent analysis.




\subsection{Response Metrics}
\label{sec:experiment:metrics}

\bstart{Completion time}
We measured task completion time as the time between dismissing the task instruction (Step 2 in \autoref{fig:task_example}) and tapping the `Done' button (Step 5 in \autoref{fig:task_example}). 
In the Animation condition, we required that participants watch the entire 12.5s loop at least once. 

\bstart{Accuracy}
We measured accuracy at two levels of granularity. 
First, we determined binary accuracy: whether or not the participant was entirely correct in their response to the task.
Second, for the five tasks that required more than one response (as indicated in \autoref{fig:tasks}), we also recorded partial correctness.

\bstart{Subjective experience}
After the final task, we asked participants to rate their overall confidence in their responses from 1 (low) to 5 (high).
We also asked them to rate the ease of using either the animated or small multiples scatterplots to complete the tasks.






\subsection{Participants}
\label{sec:experiment:recruitment}

We recruited Amazon Mechanical Turk workers with a planned sampled size of $N = 100$, assigning 50 to the Animation group and 50 to the Multiples group.
We limited our recruitment to workers in the USA with HIT approval ratings of 95\% or higher.
For consistency, we asked participants to use a phone running iOS 9 or greater or Android 5 or greater, to use either the Chrome or Safari mobile browser, and to ensure adequate battery power and a stable WiFi connection.
Considering the current federal minimum wage in the USA and the duration of our experiment (approximately 10 minutes), we paid each participant \$2 USD.
At the end of the experiment, we asked participants to copy a completion code revealed by our application into the MTurk interface.

\subsection{Implementation and Deployment}
\label{sec:experiment:implementation}

Our experimental platform is a Node.js application~\cite{Node2018} deployed as an Azure web app~\cite{Azure2018} at \url{https://aka.ms/multiples}.
We logged experimental sessions, device profiles, and responses to experimental tasks with Azure Application Insights~\cite{AppInsights2018}.
We used D3.js~\cite{Bostock2011} to implement the visualization and animation components in the client side of the application, and CSS media queries to prevent people from viewing the application from a desktop device or from a phone held in landscape mode. 
A browser cookie prevented the application from loading a second time, \revision{which we have since disabled to allow repeat visits from readers of this paper}.
Though the application was responsive to phones with varying display sizes and device-pixel ratios, our application is not user-rescalable; in other words, participants could not tap or pinch to zoom within the application. 
The source code of our application is available under an MIT open source license at \url{https://github.com/Microsoft/MobileTrendVis}.

%
%
%


\begin{figure*}[!ht]
    \small
    \centering
    {\def\arraystretch{1.3}
    \begin{tabularx}{0.925\linewidth}{XXXXXX}
        {\sffamily \tiny{\textbf{1. \highlight{Targets} \& Distractors}}} & 
        {\sffamily \tiny{\textbf{2. Instruction, \highlight{Abstract Task}}}} &
        {\sffamily \tiny{\textbf{3. Mean Completion Time (s)}}} & 
        {\sffamily \tiny{\textbf{4. Compl. Time Ratio (A / M)}}} & 
        {\sffamily \tiny{\textbf{5. Prop. Correct Responses}}} & 
        {\sffamily \tiny{\textbf{6. Diff. Prop. Correct (A - M)}}} 
        \tabularnewline
        & 
        & 
        \includegraphics[width=\linewidth]{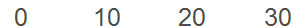} & 
        \includegraphics[width=\linewidth]{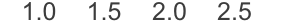} &
        \includegraphics[width=\linewidth]{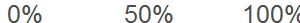} &
        \includegraphics[width=\linewidth]{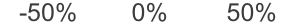}
        \tabularnewline
        \includegraphics[width=\linewidth]{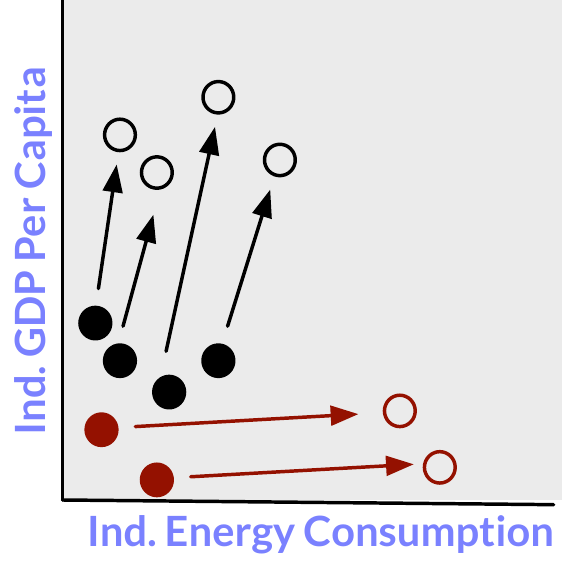} & 
        \includegraphics[width=\linewidth]{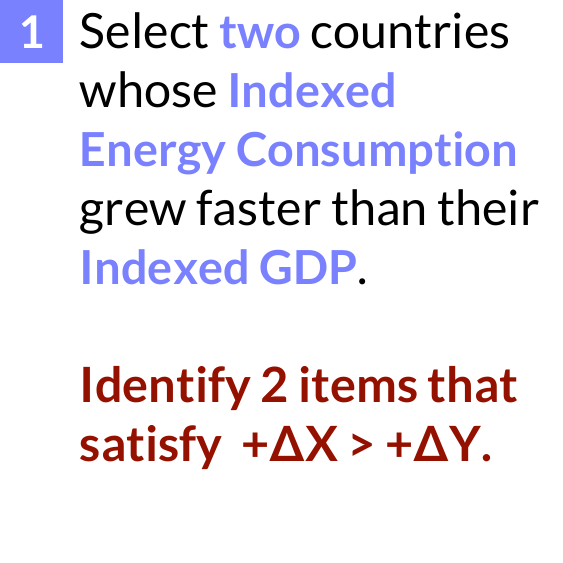} & 
        \includegraphics[width=\linewidth]{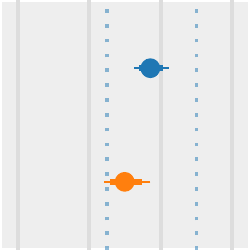} & 
        \includegraphics[width=\linewidth]{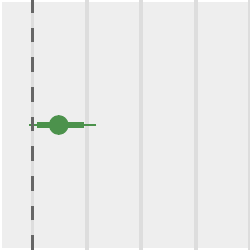} &
        \includegraphics[width=\linewidth]{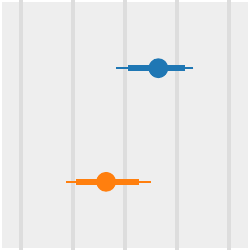} &
        \includegraphics[width=\linewidth]{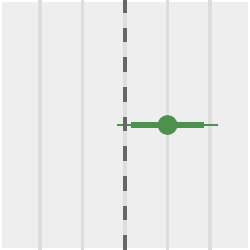}
        \tabularnewline
        \includegraphics[width=\linewidth]{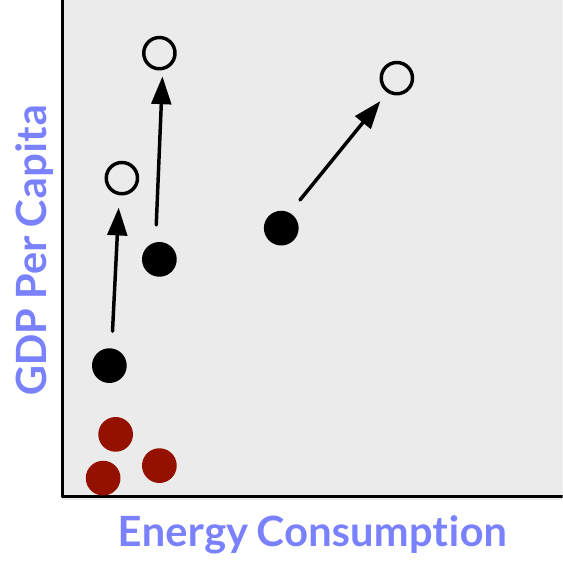} & 
        \includegraphics[width=\linewidth]{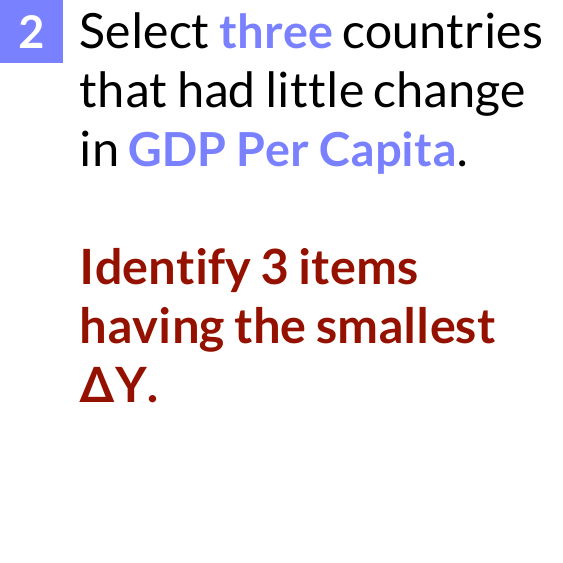} & 
        \includegraphics[width=\linewidth]{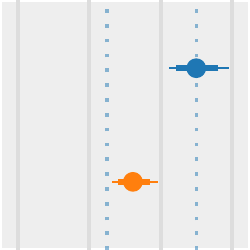} & 
        \includegraphics[width=\linewidth]{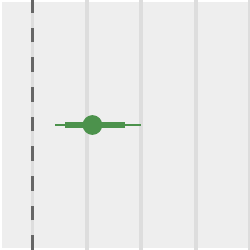} &
        \includegraphics[width=\linewidth]{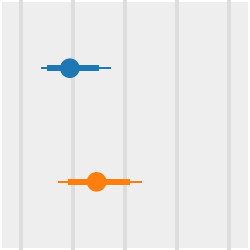} &
        \includegraphics[width=\linewidth]{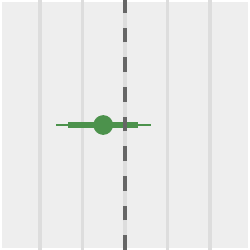}
        \tabularnewline
        \includegraphics[width=\linewidth]{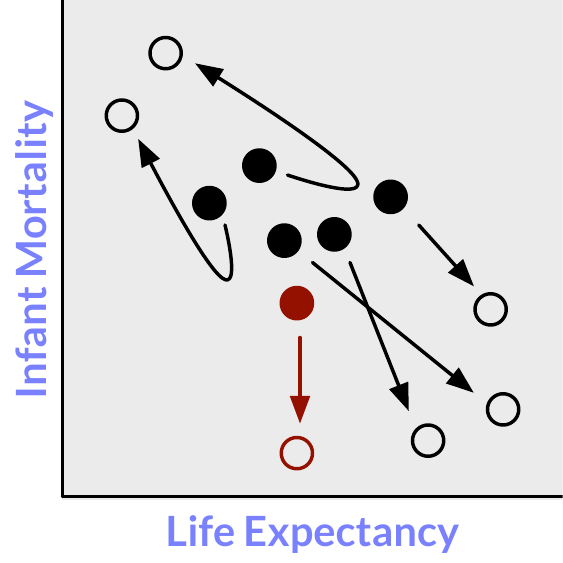} & 
        \includegraphics[width=\linewidth]{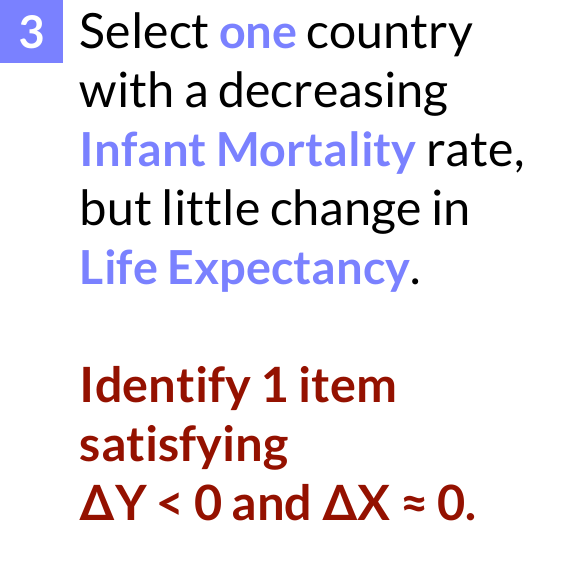} & 
        \includegraphics[width=\linewidth]{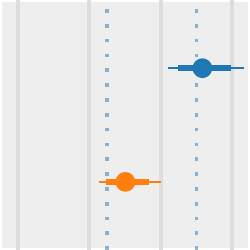} & 
        \includegraphics[width=\linewidth]{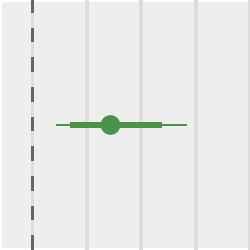} &
        \includegraphics[width=\linewidth]{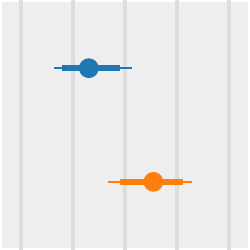} &
        \includegraphics[width=\linewidth]{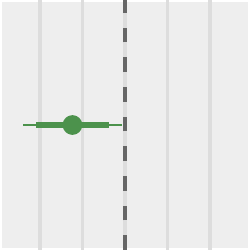}
        \tabularnewline
        \includegraphics[width=\linewidth]{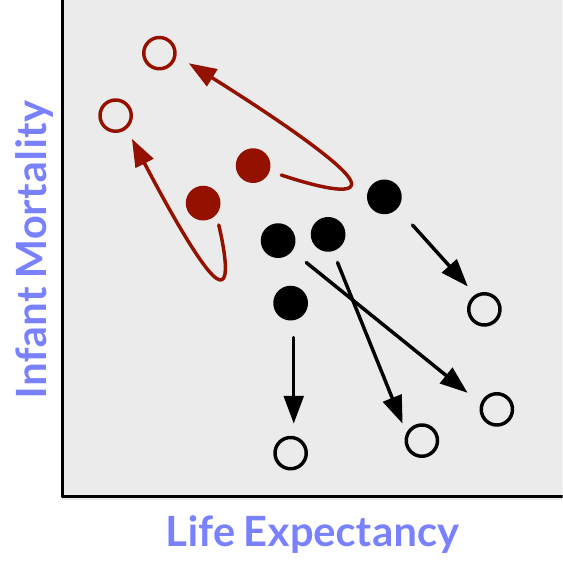} & 
        \includegraphics[width=\linewidth]{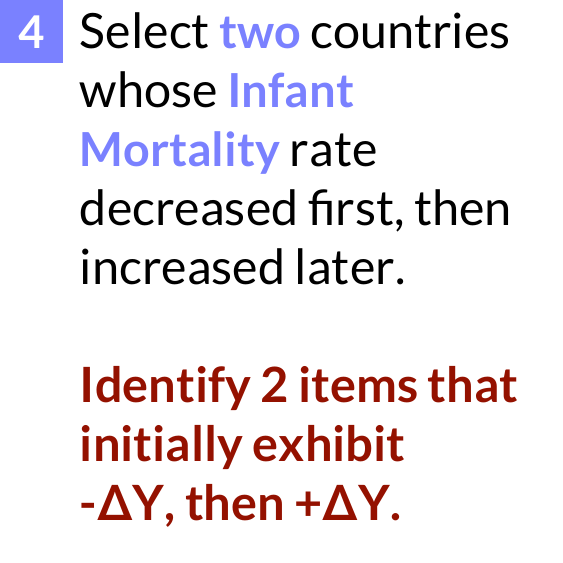} & 
        \includegraphics[width=\linewidth]{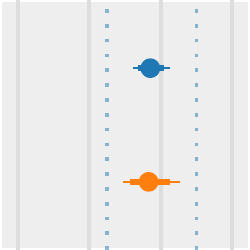} & 
        \includegraphics[width=\linewidth]{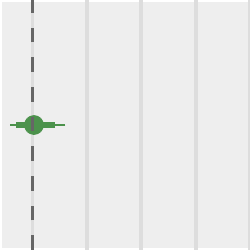} &
        \includegraphics[width=\linewidth]{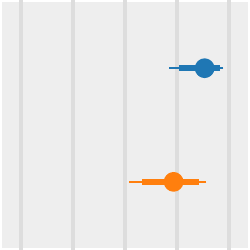} &
        \includegraphics[width=\linewidth]{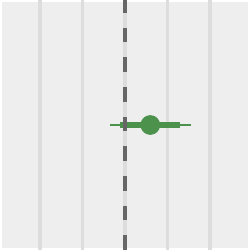}
        \tabularnewline
        \includegraphics[width=\linewidth]{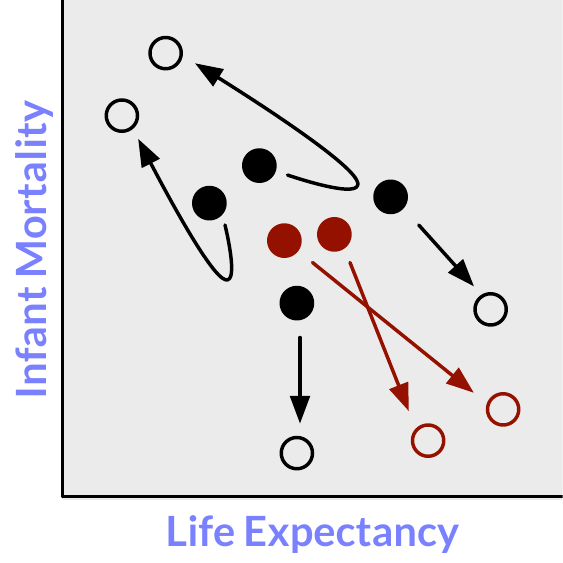} & 
        \includegraphics[width=\linewidth]{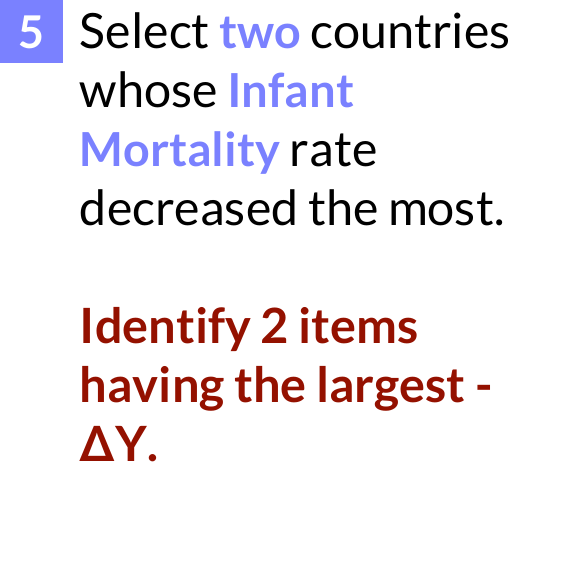} & 
        \includegraphics[width=\linewidth]{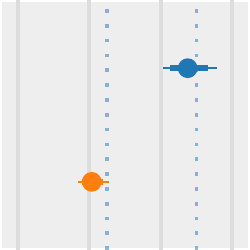} & 
        \includegraphics[width=\linewidth]{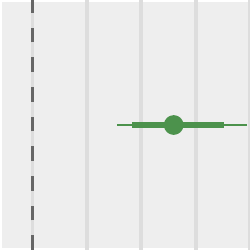} &
        \includegraphics[width=\linewidth]{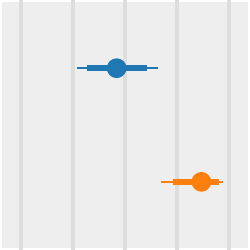} &
        \includegraphics[width=\linewidth]{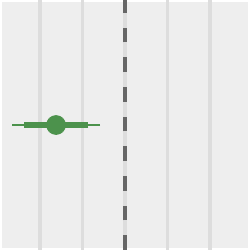}
        \tabularnewline
        \includegraphics[width=\linewidth]{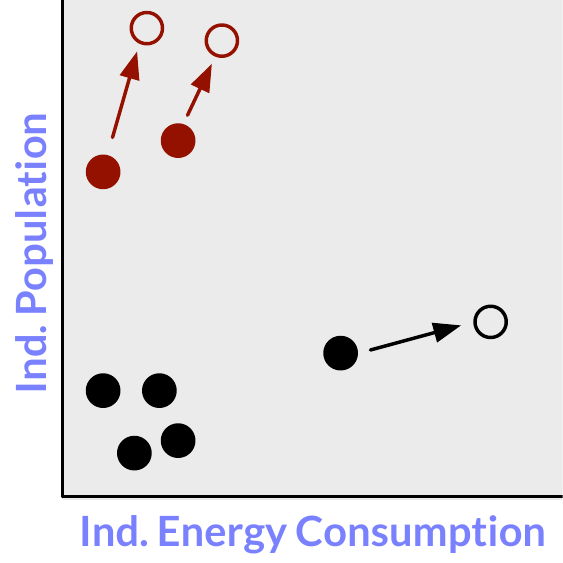} & 
        \includegraphics[width=\linewidth]{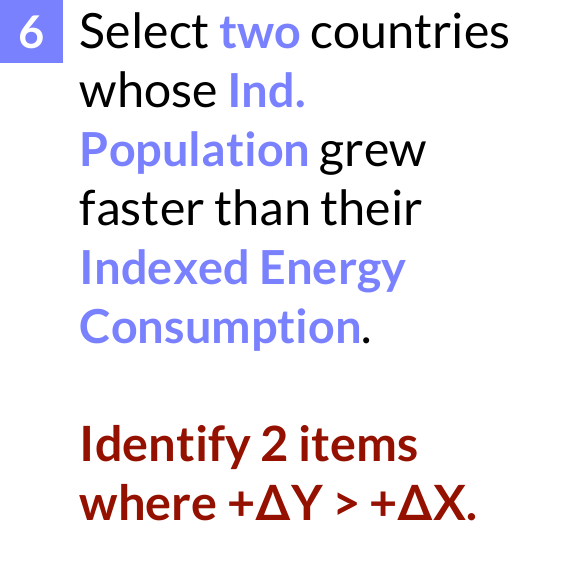} & 
        \includegraphics[width=\linewidth]{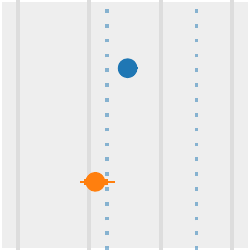} & 
        \includegraphics[width=\linewidth]{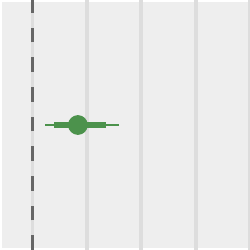} &
        \includegraphics[width=\linewidth]{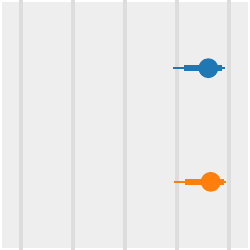} &
        \includegraphics[width=\linewidth]{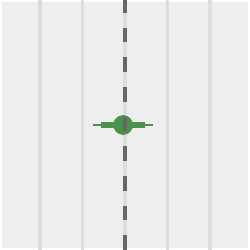}
        \tabularnewline
        \includegraphics[width=\linewidth]{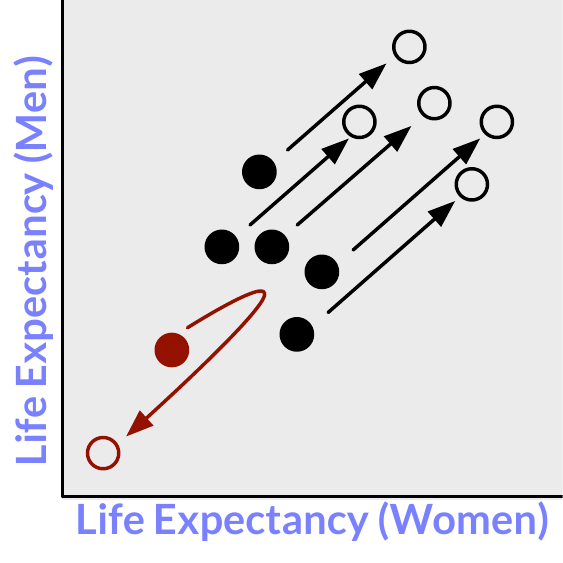} & 
        \includegraphics[width=\linewidth]{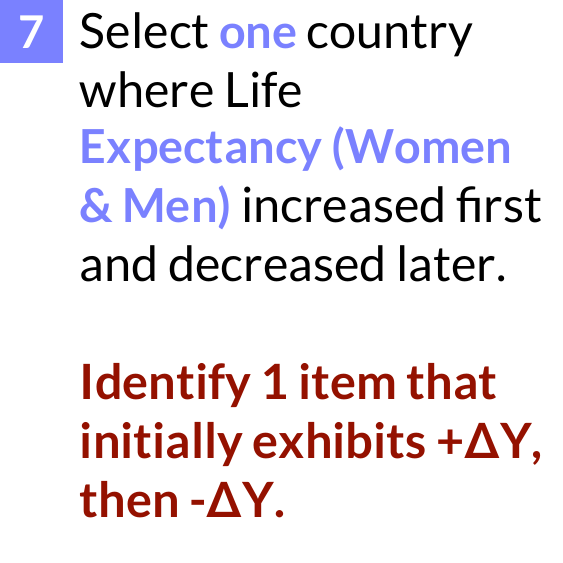} & 
        \includegraphics[width=\linewidth]{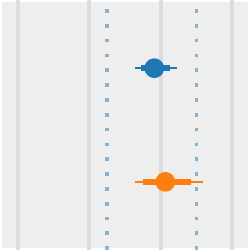} & 
        \includegraphics[width=\linewidth]{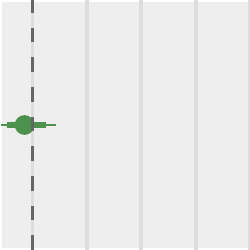} &
        \includegraphics[width=\linewidth]{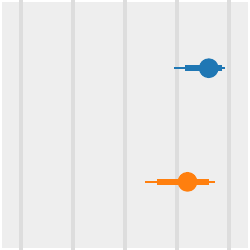} &
        \includegraphics[width=\linewidth]{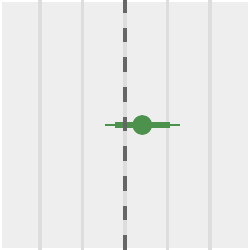}
        \tabularnewline
        \includegraphics[width=\linewidth]{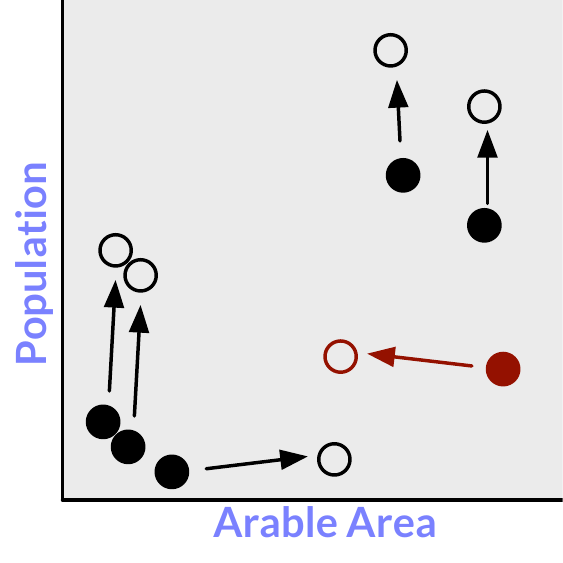} & 
        \includegraphics[width=\linewidth]{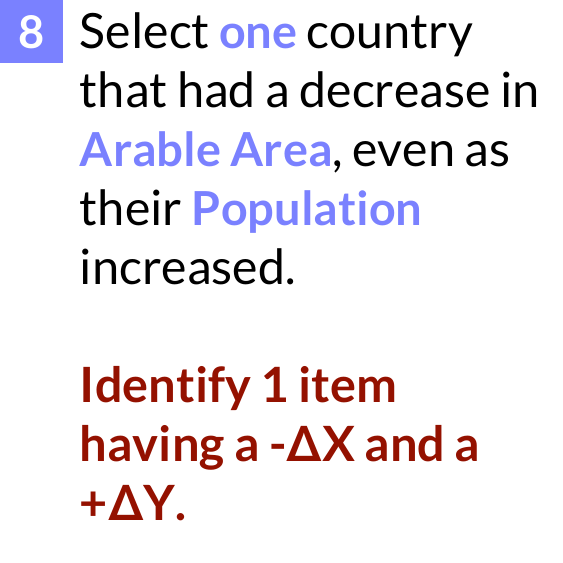} & 
        \includegraphics[width=\linewidth]{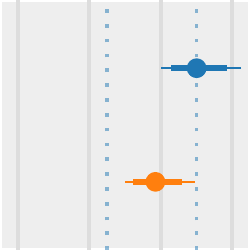} & 
        \includegraphics[width=\linewidth]{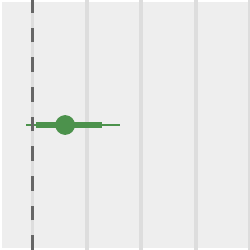} &
        \includegraphics[width=\linewidth]{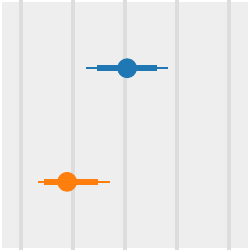} &
        \includegraphics[width=\linewidth]{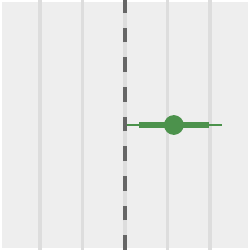}
        \tabularnewline
        \includegraphics[width=\linewidth]{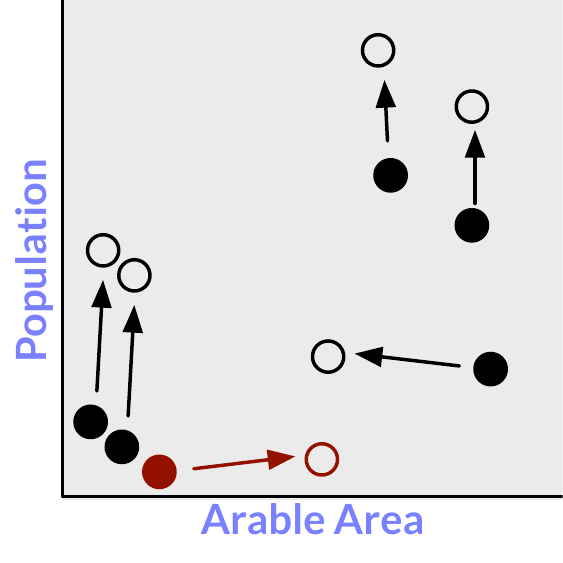} & 
        \includegraphics[width=\linewidth]{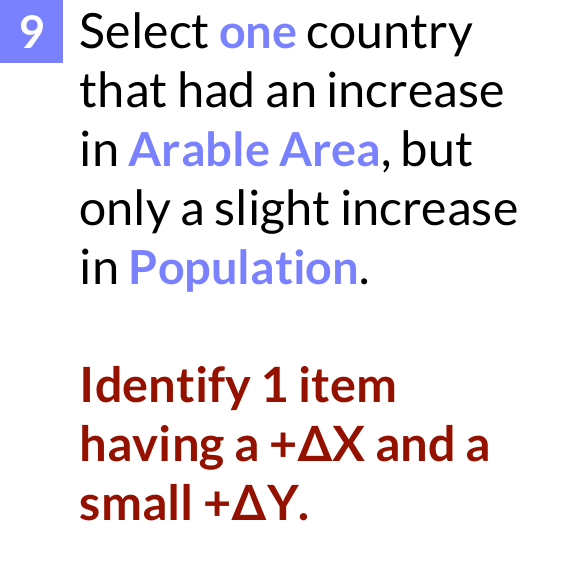} & 
        \includegraphics[width=\linewidth]{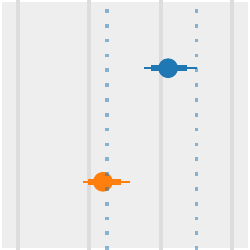} & 
        \includegraphics[width=\linewidth]{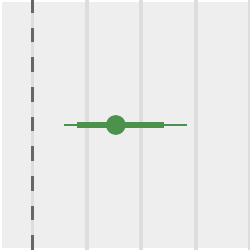} &
        \includegraphics[width=\linewidth]{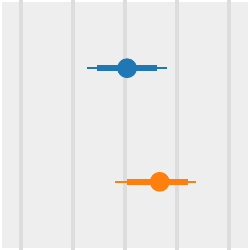} &
        \includegraphics[width=\linewidth]{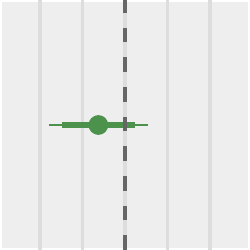}
    \end{tabularx}}
    \vspace{-0.25cm}
    \caption{The 9 tasks, mean completion times, proportions of correct responses, and estimates of effect sizes.    
      \animation{--\CIRCLE--} = \underline{A}nimation; \multiples{--\CIRCLE--} = \underline{M}ultiples; \diff{--\CIRCLE--} = Effect size estimate. Thick error bars are 95\% CIs~\cite{Krzywinski2013}, \revision{while thin error bars are conservative 99.4\% CIs that account for multiple comparisons over the 9 tasks}. \revision{Dotted blue lines in column 3 indicate 1x and 2x the length of a complete animation}. In cases where a \diff{--\CIRCLE}-- CI overlaps a dashed line (a Completion Time Ratio of 1 or a Proportional Correctness Difference of 0\%), we interpret this as inconclusive evidence for an effect~\cite{Besancon2017}.
    \label{fig:results}
    }
    \vspace{-0.5cm}
\end{figure*} 
\section{Results}
\label{sec:Results}

The total number of participants who completed our experiment was 113, however we excluded 17 participants for failing to respond correctly to our quality control question. 
We also verified that all of the participants completed the experiment using a mobile phone and a compatible browser.
Of the 96 remaining participants, 51 experienced the Animation condition and 45 experienced the Multiples condition. 
The task ordering assignments were nearly balanced among these 96 participants, with 18--21 participants assigned to each of the five task orderings; within each ordering group, the representation of the two conditions was also balanced.
Matching our expectations, the average completion time of our experiment was 9.6 minutes (\textit{SD} = 2.8 minutes).

\bpstart{\revision{Planned analyses}}
We analyze, report, and interpret our inferential statistics using interval estimation~\cite{Dragicevic2016}, and we planned all analyses in this section before collecting data.
Our collected data and analyses are provided alongside the application source code in the same \href{https://github.com/Microsoft/MobileTrendVis}{Github repository}. 
We excluded training and quality control tasks from our analysis, as well as nine trials where an interruption was detected (\ie~the application lost focus), and one additional trial where the log completion time was greater than 3 SDs from the mean log completion time for the task. 
We verified that task ordering did not have a pronounced or consistent effect on task performance by comparing the CIs of each (condition + ordering) group for each task and response metric.

\revision{Our analysis follows an approach outlined by Besan{\c c}on and Dragicevic\cite{Besancon2017}}. For each task, we report the sample mean for completion time, as well as the proportion of participants who responded correctly.
We also report 95\% confidence intervals (CIs) indicating the range of plausible values for the mean completion time and proportion of correct responses, respectively.
We also provide sample means and CIs of partial correctness for the tasks requiring multiple responses.
Lastly, we report 95\% confidence intervals for effect sizes.

\bpstart{\revision{Unplanned analyses}}
\revision{
In response to a reviewer request, \autoref{fig:results} also includes more conservative 99.4\% CIs to account for multiple comparisons across the nine tasks.}

\subsection{Completion Time}
\label{sec:Results:time}

The third column of \autoref{fig:results} shows mean completion times for each task.
We log-transform participants' completion times to correct for positive skewness and we present anti-logged geometric mean completion times~\cite{Dragicevic2016,Sauro2010}.
\revision{Though completion times vary across conditions and tasks, our participants completed most tasks in under 30s.
Note that the blue dotted lines indicate 1x and 2x the length of the animated loop (12.5s); using the length of the loop as a baseline, it is possible to spot tasks in which Animation group participants let the full loop play twice, corresponding to completion times of around 25s.} 
With the exception of Tasks 5 and 6, participants in the Multiples condition seldom completed tasks in less time than 12.5s. 

\bpstart{Participants are faster with Multiples (in most tasks)}
To compare the completion times of those in the Animation and Multiples conditions, we compute differences in log-transformed values; we present these differences as anti-logged ratios between geometric means~\cite{Besancon2017} in the fourth column of \autoref{fig:results}.
When the CI for a ratio overlaps 1.0, we see this as insufficient evidence for a difference in completion time. 
When the CI is to the left of 1.0, this is evidence for shorter completion times with Animation; in contrast, when the CI is to the right of 1.0, this is evidence for shorter completion times with Multiples.

The largest difference in task completion times occurs in Task 5: our findings indicate that this task can take up to 2.8 times longer to complete in the Animation condition than in the Multiples condition.
Tasks 3 and 9 also take longer in the Animation condition, with completions times of up to 2.2 times longer.
The gap in task completion between conditions then decreases: up to 1.9 times longer with Animation in Task 2, \revision{and up to 1.7 times longer in Task 6. 
Completion times with Animation are slightly longer in Tasks 1 and 8.}
Finally, we do not have sufficient evidence to suggest differences in completion time between the conditions in Tasks 4 or 7; we attribute the lack of a difference in these two tasks to task characteristics, which we discuss in~\autoref{sec:Results:inter}.

\subsection{Accuracy}
\label{sec:Results:accuracy}

\autoref{fig:results}'s fifth column shows the proportion of participants who responded correctly to each task, without considering partial correctness. 
We computed binomial proportions and their CIs with the PropCIs R package~\cite{Scherer2018}.
A ceiling effect is evident in Task 6, in that most participants in both conditions responded correctly. 
Otherwise, the proportion of correct responses varies considerably across conditions and tasks.

\begin{figure}[tb]
    \includegraphics[width=\linewidth]{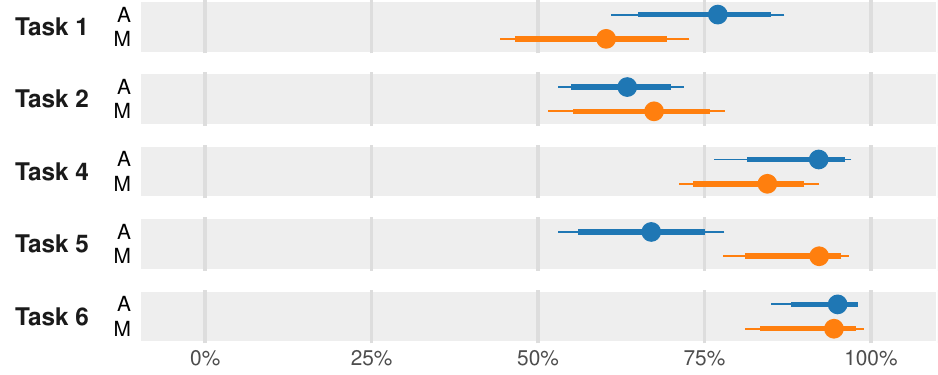}
    \caption{Mean correctness in tasks that allow for partial correctness. \animation{--\CIRCLE--} = \underline{A}nim.; \multiples{--\CIRCLE--} = \underline{M}ult. Error bars are 95\% and 99\% BCa bootstrap CIs.}
    \label{fig:partial}
\end{figure}

\autoref{fig:partial} shows the sample means and CIs of partial correctness for the tasks requiring multiple responses (Tasks 1, 2, 4, 5, and 6).
Here the CIs are BCa bootstrap confidence intervals~\cite{Besancon2017}.
The ceiling effect in Task 6 reappears here, though aside from this, the correctness varies across conditions and the remaining tasks.

\begin{figure}[tb]
    \includegraphics[width=\linewidth]{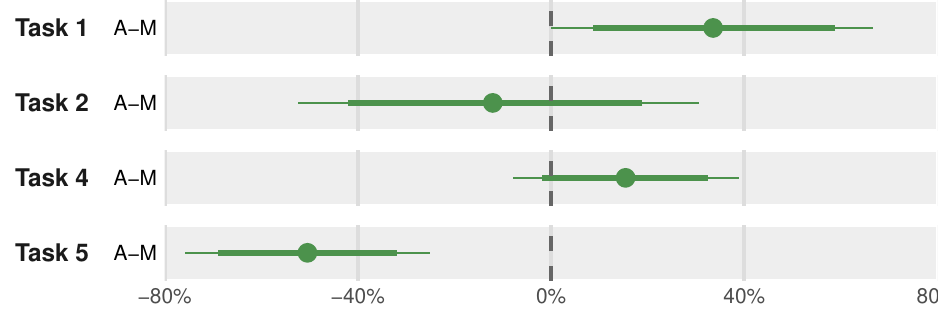}
    \caption{Effect size estimates of differences in correctness (\underline{A}nim.--\underline{M}ult.) for tasks that allow for partial correctness. Error bars are 95\% and 99\% CIs. In cases where a \diff{--\CIRCLE--} CI overlaps the dashed line (a difference of 0\%), we interpret this as inconclusive evidence for an effect~\cite{Besancon2017}.}
    \label{fig:partial-effect}
    \vspace{-0.5cm}
\end{figure}

\bpstart{Comparable accuracy with Animation and Multiples}
We report differences in proportions of correct responses in the final column of \autoref{fig:results}, as well as differences in partial correctness in \autoref{fig:partial-effect}.
In both cases, when the CI for the difference overlaps 0\%, we interpret this as insufficient evidence for a difference in accuracy. 
When the CI is to left of 0\%, this is evidence for greater accuracy with Multiples; in contrast, when the CI is to right of 0\%, this is evidence for greater accuracy with Animation.

We do not have sufficient evidence to suggest a difference in the proportion of correct responses for Tasks 2, 4, 7, and 9, nor do we find evidence for differences in partial correctness for Tasks~2 and 4. 
The remaining tasks are evenly split between Animation and Multiples.
\revision{In Task 1, those using animation are more accurate, as shown in \autoref{fig:partial-effect}. 
In Task 8, the proportion of correct responses among people who use animation would be up to 49\% higher than the proportion of correct responses among people who use small multiples.}
\revision{Meanwhile, the proportion of correct responses among those using small multiples would be up to 52\% higher than the proportion of correct responses among those using animation for Task~3 and up to 60\% higher for Task 5} (and those using small multiples would be up to 69\% more accurate, as shown in \autoref{fig:partial-effect}).



\subsection{Subjective Experience}
\label{sec:Results:subj}

\begin{figure}[h!]
    \includegraphics[width=\linewidth]{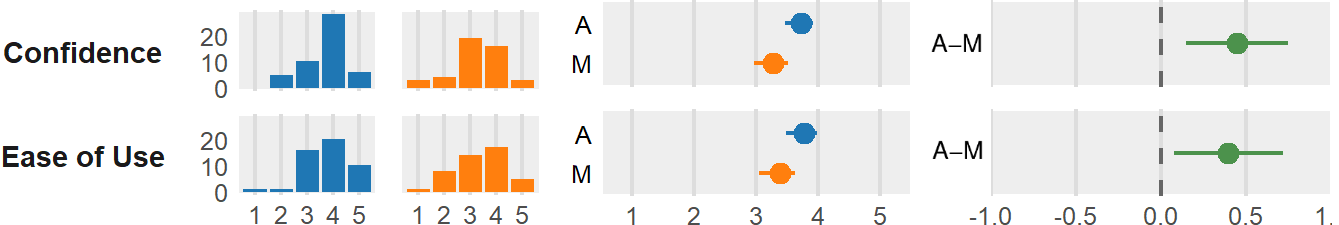}
    \vspace{-0.4cm}
    \caption{Participants' subjective experience ratings from 1 (low) to 5 (high).}
    \label{fig:subj}
    \vspace{-0.15cm}
\end{figure}

\autoref{fig:subj} shows participants' mean ratings along with BCa bootstrap confidence intervals on a scale from 1 (low) to 5 (high) with respect to their confidence in their responses as well as their perceived ease of using the charts to respond to the tasks.

\bpstart{Participants are slightly more confident with Animation}
Those in the Animation condition report feeling more confident in their task responses (up to 0.76 points higher on a 1--5 scale) than those in the Multiples condition (\autoref{fig:subj}, top right).
When asked about whether the charts they used were easy to use in terms of responding to the task instructions, those using animation provide higher ratings (up to 0.71 points higher, \autoref{fig:subj}, bottom right). 


\subsection{Results Interpretation}
\label{sec:Results:inter}

The participants who used a small multiples variant of a scatterplot completed seven out of nine trend comparison tasks more quickly than the participants who used an animated variant. 
This finding is similar to Robertson~\etal's results~\cite{Robertson2008}, where participants were faster to complete tasks with a small multiples variant of a scatterplot than with an animated variant; however, their animation was slightly shorter (10s) and they permitted participants to interactively pause or scrub through the animation. 
Their experiment also involved both large and small datasets, and since they did not disclose completion time results specific to their small dataset condition, we cannot directly compare our results.

With respect to accuracy, small multiples do not appear to have a substantial advantage, which is unlike what Robertson~\etal~found~\cite{Robertson2008}, where those using small multiples were more accurate. 
In our case, we found no evidence for a difference in the proportions of correct responses or differences in accuracy for five of the nine tasks.
In Tasks 3 and 5, a higher proportion of participants responded correctly in the Multiples condition; however, the converse was true in Tasks 1 and 8.

We conclude that small multiples remain to be a viable design choice even for small mobile phone displays, as we found that they are associated with faster trend comparisons and that these judgments are comparably accurate to those made with an animated scatterplot.
Therefore, the ``small multiples on desktop, animation on the phone'' design pattern~\cite{Boyer2015} may be unjustified for scatterplots, at least with a similar (or fewer) number of grid cells relative to what we used in our experiment. 
Despite this general pattern, we observed a possible trade-off between completion time and error in Tasks 1 and 8, where participants in the small multiples condition were faster but less accurate, which leads us to our discussion of task-specific characteristics.




To interpret the results for individual tasks, we characterize them in terms of their number of required responses and the trajectories of their target and distractor items (see the \textbf{\highlight{abstract task descriptions}} in the second column of \autoref{fig:results}).
To compare the animated trajectories beyond the diagrams shown in \autoref{fig:results}, we also provide a supplemental video depicting all tasks performed in both conditions.

\lettrine[lines=5,findent=2mm,nindent=-.5mm]{\includegraphics[width=15mm]{figures/T4.pdf}\includegraphics[width=15mm]{figures/T7.pdf}} 
{\bf Self-occluding} \textbf{trails} \textbf{and} \textbf{common} \textbf{fate}.
Tasks 4 and 7 share an interesting commonality: both asked participants to select targets that reverse direction in $\Delta${\tt y}: a decrease followed by increase in the case of Task 4 and the converse in Task 7. 
\autoref{fig:teaser} shows instances of the stimuli used for Task 4 in both conditions, where the correct responses are L + M in the Animation condition and F + N in the Multiples condition; recall that item labels and colors are randomly assigned and are not consistent between these two instances.
Reversals in {\tt x} or {\tt y} form loops in connected scatterplots~\cite{Haroz2016}, and depending on the behavior of the other variable, a loop can occlude itself, and thus a loop in the ``trail'' of our small multiples variant of a scatterplot may be less salient.
With an animated scatterplot, the Gestalt principle of {\it common fate}~\cite{Palmer1999} would suggest that the distractor items in these tasks are likely to be perceptually grouped because they move cohesively in one direction.
Meanwhile, the target items initially travel in the same direction as the distractors, but later move in a markedly different direction.  
This behavior may explain our lack of evidence for a difference in performance between small multiples and animation in these two tasks.

\begin{figure}[h!]
    \animategraphics[width=0.5\columnwidth,loop,autoplay]{10}{figures/task1/task1_}{005}{129}
    \includegraphics[width=0.5\columnwidth]{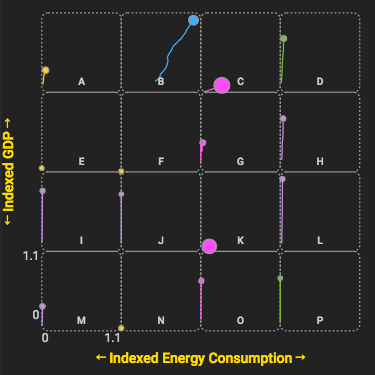}
    \caption{Task 1 asks participants to identify items where $\Delta${\tt x} $>$ $\Delta${\tt y}. Left: the Animation condition (correct responses: H and O; \textbf{open this PDF in Acrobat Reader to view the animation}). Right: Multiples condition (correct responses: C and K). Note that item labels and colors are randomly assigned and are not consistent between these two instances.} 
    \label{fig:task1}
    \vspace{-0.2cm}
\end{figure}

\lettrine[lines=5,findent=2mm,nindent=-.5mm]{\includegraphics[width=15mm]{figures/T1.pdf}\includegraphics[width=15mm]{figures/T8.pdf}\includegraphics[width=15mm]{figures/T9.pdf}} 
{\bf Outliers} \textbf{and} \textbf{shared} \textbf{coordinate} \textbf{spaces}.
Tasks 1, 8, and 9 involve comparisons that take into account both $\Delta${\tt x} and $\Delta${\tt y}.
\revision{In Task 8, the Animation condition participants were more accurate in their responses, while in Task 1, they were slightly more accurate;} \autoref{fig:task1} shows instances of the stimuli used for Task 1 in both conditions.
\revision{One interpretation of this finding is that the target items in these tasks start or end in {\tt x-y} positions that are quite far from the distractor points}, meaning that the targets are position outliers during some or all of the time series.
Trend comparisons that consider both $\Delta${\tt x} and $\Delta${\tt y} may benefit from a shared coordinate space; this is provided by our Animation condition, whereas individual items are plotted within their own grid cell in our Multiples condition.

Veras and Collins~\cite{Veras2019} recently studied outliers in 2-state animated scatterplots, where points may be outliers in terms of speed and direction of motion.
They also found that different visual channels affect outlier saliency to varying extents, and as a result it can be difficult to consistently detect them in animated scatterplots.
In Tasks 1 and 8, the target items can be described as being outliers in terms of direction of motion.
Other dimensions such as their speed of motion, their final position, their size, or their color may have also contributed to their higher saliency.
The speed cue is inherently absent in the Multiples condition, and thus the participants needed to rely on the remaining cues, thereby resulting in lower accuracy relative to the performance of those in the Animation condition.

\lettrine[lines=5,findent=2mm,nindent=-.5mm]{\includegraphics[width=15mm]{figures/T3.pdf}\includegraphics[width=15mm]{figures/T5.pdf}} 
{\bf Identifying} \textbf{straight} \textbf{or} \textbf{long} \textbf{trails}.
Tasks 3 and 5 involve comparisons that take into account change in only one direction ($\Delta${\tt y} in both cases).
For both of these tasks, the Multiples condition participants were more accurate in their responses.
\autoref{fig:teaser} shows instances of the stimuli used for Tasks 3 and 5. 
Tasks 3 and 5 share the same pair of {\tt x} and {\tt y} axes, and as mentioned in \autoref{sec:experiment:design}, we shuffled the order of task presentation such that no two consecutive tasks shared the same axis pairing. 
Participants needed to identify a straight vertical trail in Task 3 and the two trails with the greatest $\Delta${\tt y} in Task 5. 
Referring again to \autoref{fig:teaser}, the correct response for Task 3 is A in the Animation condition and M in the Multiples condition; in Task 5, the correct responses are F + P in the Animation condition and E + G in the Multiples condition.
Making distance and angle judgments such as these are difficult in the context of animation, as there are no trails with which to compare lengths, and thus these judgments must rely on memory rather than eye movements, recalling Munzner's {\it eyes beat memory} maxim~\cite{Munzner2014}.


\section{Discussion}
\label{sec:discussion}

Based on our study results and interpretations, we offer implications for visualization design on mobile devices and discuss potential extensions to this work. We then reflect on our evaluation methodology. 


\revision{We preface this discussion with a brief comment on the generality of the stimuli used in our experiment. 
While the trajectories of target and distractor items in our tasks are not representative of all possible trend scenarios, we expect them to be representative of scenarios such as those seen in Rosling's TED conference presentations~\cite{Rosling2006,Rosling2007}.
These scenarios include trajectory reversals, trajectories exhibiting impressive gains or reductions along a single dimension, or outlier trajectories with respect to their direction of motion.
In practice, the orientations of these trajectories will differ from those seen in our nine experimental tasks.
Despite this, we do not have any reason to suspect that our participants' performance was affected by specific trajectory orientations.}

\subsection{Implications for Design}
\label{sec:discussion:design}

When visualizing a multivariate dataset to support trend comparison tasks on a mobile phone display, we advocate for the use of a static small multiples scatterplot with trajectories encoded as trails over an animated scatterplot.   
However, there may be exceptions: if the viewer is expected to identify trajectory reversals, small multiples may not maintain its advantage.
Similarly, if the viewer is expected to notice outliers, animation within a shared {\tt x-y} coordinate space may be preferable to small multiples.
Another option is a single unified trails chart (a multi-series connected scatterplot), 
which may support outlier detection to the same extent as animation.
However, such a design 
requires plotting all items within a shared {\tt x-y} coordinate space; the drawback of which is increased visual clutter incurred by plotting all of the item trails within this small space, which we observed in our early prototyping.


Our Animation participants reported feeling slightly more confident, and they also reported a higher ease of use. 
A compromise design solution would be to incorporate both animation and small multiples.
During our prototyping, we considered but ultimately abandoned an animated small multiples design.
We found that this design required tracking too many moving items within their individual coordinate spaces. 
Instead, we are more optimistic about sequencing animation and small multiples.
The former could be used as a means to introduce the data and emphasize that it changes over time.
\revision{Following the completion of the animation, a transition to a static small multiples scatterplot could then allow the viewer to make accurate judgments about trajectory length or angle}.
Boyandin~\etal~\cite{Boyandin2012} made a similar recommendation following their qualitative study of animation and small multiples in the context of flow maps, in which they noted how their participants used the two design alternatives to perform different tasks.

Regardless of whether a designer uses animation or small multiples to support trend comparison tasks, designers should consider techniques for highlighting or annotating instances of trajectory reversals and outliers, as the tasks that involved these patterns were ones in which we saw differences in performance between the two design variants. 
We take inspiration from Bryan~\etal's automatic annotation of features in area charts~\cite{Bryan2017}, in that trajectory reversals and outliers in animated or small multiples scatterplots could be automatically highlighted or annotated.
However, designers will need to consider the limitations of display size and the saliency of data items relative to the saliency of highlights and annotations. 

\subsection{Potential Extensions to this Study}
\label{sec:discussion:extensions}


Each of our tasks involved scatterplots with 16 data items and 25 time steps, with a smoothly interpreted animation lasting 12.5 seconds, or 500ms per time step. 
These parameters were inspired by the small dataset condition in Robertson~\etal's experiment, which was in turn informed by Rosling's presentations.
We now consider the implications of varying these parameters.

\bpstart{Varying the number of items}
We are skeptical about the viability of animation or small multiples on mobile phones in the context of larger datasets, \revision{such as the dataset of 80 nations used in Robertson~\etal's experiment}.
This skepticism is based on \revision{the physical dimensions of mobile phone displays and our initial experimentation with different dataset sizes.
A larger dataset will result in increased occlusion in an animation design or smaller grid cells and items in a small multiples design. 
For example, consider a 9x9 small multiple grid akin to the one shown in \autoref{fig:robertson}-left; a single grid cell would only be 35pt wide ($\sim$0.3cm) on an iPhone SE when held in portrait mode.}
We predict that a small multiples design for datasets with fewer than 16 items will be superior to animation, as larger small multiple grid cells provide a greater resolution with which to plot items and their trails. 

\bpstart{Varying the number of time steps and their pacing}
Our results indicate that participants in the Multiples condition seldom completed a task in less time than it took those in the Animation condition to watch a single loop of the animation, which suggests that the length of our animation was not too fast considering the number of time steps in the data (see \autoref{fig:results}, third column).  
If we increase the number of time steps, the trajectories in our small multiples design may become more difficult to distinguish~\cite{Haroz2016} due to occlusion, particularly when trajectory reversals occur.
Similarly, if we add time steps without altering the pace of the animation, a viewer may have considerable difficulty making comparisons that rely on memory over the course of a long animation.
If we instead adjust the pace of the animation and maintain the length of the animation, the result may be excessively jittery and tracking individual items is likely to become increasingly difficult~\cite{Alvarez2007}.
Another alternative would involve staggering the animation, pausing briefly at each time point (\eg~\cite{PewGiF2012}).
Altogether, varying parameters relating to time should be examined in a series of future experiments. 

\bpstart{Focusing on reversals and outliers}
\revision{We envision further experimentation with design alternatives that allow people to identify trajectory reversals and outliers. 
This will likely require a systematic generation of synthetic stimuli modeled after the reversals and outliers apparent in the real United Nations data that we used in our current experiment.}
Generating synthetic stimuli would also permit a finer control over other visual variables such as color and size, as Veras and Collins' recently showed that these and other visual variables contribute unevenly to outlier detection~\cite{Veras2019}. 

\bpstart{Other chart types}
Additional research is needed to determine if our results generalize beyond scatterplots to other visual representations of time-varying data.
These might include charts depicting seasonal temperature change over time (\eg~\cite{Hawkins2016}), graphs depicting the evolution of a node-link network~\cite{Archambault2011,Bach2014}, dynamic flow maps~\cite{Boyandin2012}, or animated symbol maps depicting the spread of entities over space and time (\eg~\cite{NPRWalmart}).

\bpstart{Interactive alternatives}
In our experiment, we focused on trend comparisons in the absence of interaction.
It is nevertheless worth considering simple interactions beyond a scrubbable playhead for controlling the rate and current frame of the animation. 
For instance, the viewer could interactively toggle between a small multiples view and a ``large single'' view of a single time step or of a single data item~\cite{vandenElzen2013}, such as by using a small multiples view as an interactive table of contents.
A sequence of small multiples and single scatterplots could also be curated in such a way that the viewer only has to swipe between them, similar to how people swipe to navigate card-based stories in social media applications.

\subsection{Methodological Reflection}
\label{sec:discussion:method}

Our experiment and its results should be viewed as a first step toward assessing the viability of animation and small multiples for trend visualization on mobile phones. 
We encourage mobile visualization designers to perform their own evaluation of these design alternatives and to share details regarding their reception among the people that use them.

Experiments such as ours inherently involve a trade-off between the control over potential confounds and external validity.
In our case, we were fortunate to recruit many participants via a crowdsourcing platform and they performed the experiment using their own mobile phone.
Yet, crowdsourced experiments also have drawbacks for visualization research~\cite{Borgo2017,Borgo2018}. 
While the absence of a co-located experimenter to directly observe participants may elicit behavior that is more representative of typical performance, we could not control the contexts in which participants performed our experiment and the potential confounding effects of distractions, interruptions, and screen size.

\bpstart{Ensuring response quality}
In light of threats to validity, we refined our experimental protocol and our analysis over the course of three pilot experiments.
In our first pilot experiment, we recruited 20 current and former colleagues for a pilot experiment.
One of the authors observed four in-person pilot sessions, while the remaining pilot participants were remote; in both cases, we received feedback with respect to the usability of the experimental application and the relative difficulty of the tasks. 
Anticipating the threat of distractions and interruptions, we implemented a time-out prompt and a corresponding log event that would appear in cases when a task remained incomplete after a minute of inactivity. 
We were also able to detect and log instances in which the mobile browser tab loses focus due to an interruption.
Logging these events allowed us to exclude excessively long or interrupted trials from our subsequent data analysis.

In our second and third pilot experiments, we recruited 16 participants via Mechanical Turk. 
In the second pilot experiment, we noted a pronounced drop in accuracy as well as a pronounced reduction in task completion time relative to our first pilot experiment.
This prompted us to refine our quality control task and add a minimum instruction reading time of 5 seconds preceding each task (see \autoref{fig:task_example}, Step 1).
\revision{We also noted an imbalance in the allocation of conditions and shuffled task orderings across participants.
To address this imbalance, we implemented a server-side condition rebalancing protocol that kept track of the experiment's progress in terms of which participants had completed, which participants had yet to complete, and which had closed the application without completing.}
These two improvements further distinguish this experiment's design from that of our previous crowdsourced mobile visualization experiment~\cite{Brehmer2019}, which employed neither a mandatory instruction reading time or a condition rebalancing protocol.
The results of our third pilot experiment indicated that our quality control measures resulted in higher overall response quality as well as more even allocation of conditions and task orderings.

Researchers should continue to investigate techniques for ensuring instruction compliance and response quality from a crowd worker participant using a mobile phone. 
For instance, by requesting access to the phone's orientation, microphone, WiFi, and ambient light sensors, we might be able to detect, discard, and replace trials affected by a potentially distracting change in the environment.

\bpstart{Toward a mobile visualization evaluation platform}
We see this work as another step toward establishing and refining a mobile visualization evaluation methodology, one that incorporates and extends crowdsourced protocol used in previous work, \revision{such as in our recent experiment on alternative displays of ranges over time~\cite{Brehmer2019} or in Schwab~\etal's recent experiment on pan and zoom interactions across mobile and desktop devices~\cite{Schwab2019}}. 
Such a methodology will allow the visualization community to establish best practices and guidelines for visualizing data on mobile phones, which are undoubtedly needed as the prevalence of mobile-first and mobile-only data visualization continues to grow.

In conjunction with methodological development, the research community would also benefit from platforms and tools for performing mobile visualization studies. 
The open-source experimental applications used in this study and in our previous study~\cite{Brehmer2019} share a common architecture, one which could be repackaged as a modular application framework or as a mobile-specific extension to existing visualization experimentation platforms such as Experimentr~\cite{Harrison2019}. 
\revision{However, moving beyond visual perception experiments to encompass experiments involving the comparison of alternative interaction techniques will require additional effort on behalf of framework developers, so as to ensure both compatibility across devices and participant response quality.}


\section{Conclusion}
\label{sec:conclusion}

Motivated by a growing practical interest in visualization design for mobile applications, we conducted a crowdsourced experiment to compare the efficacy of animated and small multiples variants of scatterplots on mobile phones for supporting trend comparison tasks.
This work is in many ways a sequel to Robertson~\etal's experiment~\cite{Robertson2008}, which examined similar design choices in the context of larger displays.

In our experiment, 96 participants completed nine trend comparison tasks.
We found that participants using a small multiples variant of a scatterplot consistently performed the tasks in less time than those using an animated variant.
The accuracy results were more task-dependent, as some tasks appeared to favor those using small multiples, while others appeared to favor those using animation.
By examining the characteristics of the nine different tasks, we offered interpretations of these results, determining that tasks involving length and angle judgments may be better supported by small multiples, while animation may be better suited for supporting judgments about trajectory reversals and directional outliers.
We also noted the continued subjective appeal of animation, and thus in the absence of a clear accuracy advantage over small multiples, animation remains to be a viable design choice for some tasks.
We hope mobile visualization designers can benefit not only from our study results but also from the evaluation methodology we have been striving to refine and share.

\acknowledgments{We thank Pierre Dragicevic, Steve Haroz, and Ed Cutrell for their help with our study design and \revision{our statistical analysis}. We also thank Roland Fernandez for sharing material from the 2008 study~\cite{Robertson2008}. Finally, we thank our pilot participants for their time and effort.}

\pagebreak

\bibliographystyle{abbrv-doi}

\bibliography{main}
\end{document}